\documentclass[hyper]{JHEP3}

\usepackage[dvips]{graphicx}
\usepackage{epsfig,multicol,amssymb,amsmath,cancel}

\newcommand\fverb{\setbox\fverbbox=\hbox\bgroup\verb}
\newcommand\fverbdo{\egroup\medskip\noindent%
            \fbox{\unhbox\fverbbox}\ }
\newcommand\fverbit{\egroup\item[\fbox{\unhbox\fverbbox}]}
\newbox\fverbbox

\def\bea{\begin{eqnarray}}
\def\eea{\end{eqnarray}}
\def\beq{\begin{equation}}
\def\eeq{\end{equation}}

\title{Peccei-Quinn invariant extension of the NMSSM}

\author{Kwang Sik Jeong, Yutaro Shoji, Masahiro Yamaguchi \\
Department of Physics, Tohoku University, Sendai 980-8578, Japan  \\
E-mail :
\email{ksjeong@tuhep.phys.tohoku.ac.jp},
\email{yshoji@tuhep.phys.tohoku.ac.jp},
\email{yama@tuhep.phys.tohoku.ac.jp} }

\preprint{TU-895}

\abstract{
We study a Peccei-Quinn invariant extension of the next-to-minimal
supersymmetric Standard Model (NMSSM), which
turns out to be free from the tadpole and domain wall problems.
Having a non-renormalizable coupling to the axion superfield,
the SM singlet added to the Higgs sector can naturally generate
an effective Higgs $\mu$ term around the weak scale.
In the model, the lightest neutralino is dominated by the singlino,
which gets a mass only through mixing with the neutral Higgsinos.
We explore the phenomenological consequences resulting from the
existence of such a relatively light neutralino.
The coupling of the SM singlet to the Higgs doublets is constrained
by the experimental bound on the invisible $Z$-boson decay width.
Under this constraint, we examine the properties of the SM-like Higgs
boson paying attention to its mass and decays.
We also demonstrate a UV completion of the model in SU(5) grand
unified theory with a missing-partner mechanism.
}


\begin{document}

\section{Introduction}

The next-to-minimal supersymmetric Standard Model (NMSSM) introduces
a SM singlet $S$ to explain the origin of a supersymmetric Higgs $\mu$
term of the MSSM \cite{Review-NMSSM}.
However, if $S$ is a true singlet under all symmetries, it becomes
difficult to embed the NMSSM into a more fundamental theory
such as a grand unified theory (GUT).
This is because the GUT partners of the MSSM Higgs doublets
also couple to $S$ and radiatively generate a large tadpole for $S$,
destabilizing the gauge hierarchy \cite{Tadpole-problem1}.
Non-renormalizable interactions are another source of large tadpoles
\cite{Tadpole-problem2}.
A symmetry under which $S$ transforms non-trivially can solve the
tadpole problem, but generally introduces another problem.
If one considers a discrete symmetry, dangerous domain walls would
be formed in the early universe \cite{Domain-wall-problem,DWP}.
On the other hand, a global symmetry spontaneously broken by $S$
and the Higgs doublets would give rise to an unacceptable visible
axion.
Additional structure is thus needed to generate $\mu$ dynamically
from the coupling of $S$ to the Higgs doublets while providing
a viable framework for the grand unification.

In this paper, we point out that the difficulties arising due to
the singlet $S$ can be avoided in a Peccei-Quinn invariant extension
of the NMSSM (PQ-NMSSM) where the PQ symmetry is spontaneously broken
at a scale much higher than the weak scale by the axion superfield.
The PQ symmetry forbids the generation of large tadpoles for $S$
while solving the strong CP problem \cite{PQ-mechanism}.
Furthermore, the domain wall problem can be resolved in the presence
of PQ messengers that couple to the axion superfield.
It also turns out that a non-renormalizable coupling of $S$ to
the axion superfield naturally leads $S$ to get a vacuum expectation
value around the weak scale.

The Higgs and neutralino sectors are considerably modified by the addition
of $S$.
In the PQ-NMSSM, the lightest neutralino consists mostly of the singlino
because it acquires a small mass through mixing with the neutral Higgsinos
after the electroweak symmetry breaking.
The presence of such a relatively light neutralino leads to
phenomenological consequences different from other NMSSM models.
In particular, the LEP bound on the invisible $Z$-boson decay width
places a stringent constrain on the coupling of $S$ to the Higgs
doublets if the decay mode is kinematically allowed.
This constraint becomes important at large $\tan\beta$.
We also note that loops involving the Yukawa coupling of the singlino
give an additional positive contribution to the SM-like Higgs mass.
This contribution is insensitive to $\tan\beta$, and arises when the
Higgsinos are lighter than other MSSM sparticles.
Another interesting feature is that the decay of the SM-like Higgs boson
into a pair of the lightest neutralino can be dominant at low $\tan\beta$.

Since interactions of $S$ are controlled by the PQ symmetry, it is
possible to embed the PQ-NMSSM into GUT models without the tadpole problem.
The associated UV completion is then related to the doublet-triplet splitting
problem.
We find that incorporating the PQ symmetry in a missing-partner
model for supersymmetric SU(5) GUT \cite{Missing-partner,PQ-missing-partner}
can lead to the PQ-NMSSM at low energy scales.
In addition, it naturally achieves a phenomenologically acceptable value
of the axion decay constant as $F_a \sim \sqrt{M_{\rm SUSY}M_{Pl}}$ with
$M_{\rm SUSY}$ being the SUSY breaking scale.
The PQ symmetry is also important for suppressing harmful dimension 5
operators for proton decays.

This paper is organized as follows.
In the next section, we present the model and discuss its general
properties and various constraints on the singlet couplings.
Then, in section 3, we construct a low energy effective theory below the SUSY
breaking scale to examine the phenomenological aspects resulting
from the existence of a singlino-like light neutralino.
Section 4 is for the discussion on how to UV complete the model.
The PQ-NMSSM can arise as a low energy theory of a missing-partner GUT model.
Section 5 is the conclusion.

\section{PQ-invariant extension of the NMSSM}

In this section, we extend the NMSSM\footnote{
Extensions of the MSSM with a SM singlet have been received revived attention
\cite{Recent-NMSSM1,Nakayama,Recent-NMSSM2,Recent-NMSSM3,Recent-NMSSM4}
after the first results on the Higgs search at the LHC had been announced.
}
to incorporate the PQ symmetry and study the properties of the model.
The PQ-invariant extension turns out not only to provide
a solution to the strong CP problem \cite{Kim:2008hd} but also to solve
both the tadpole and domain wall problems.
We also examine constraints on the coupling of $S$ to the Higgs doublets.

\subsection{Model}

In the PQ-NMSSM, an effective Higgs $\mu$ term is generated by the vacuum
expectation value of $S$ which couples to the Higgs doublets and to the axion
superfield $X$ through the PQ-invariant interactions
\bea
\label{PQ-NMSSM}
{\cal L} =
\int d^2 \theta \lambda SH_uH_d
+ \int d^4\theta\, \kappa \frac{X^{*2}}{M_{Pl}}S + {\rm h.c.},
\eea
for $(X,S,H_u,H_d)$ carrying the PQ charges as $(1,2,-1,-1)$.
All other terms involving $S$ are forbidden by the PQ symmetry in the
renormalizable superpotential.
A superpotential term $X^2H_uH_d/M_{Pl}$, which is allowed by $U(1)_{\rm PQ}$,
can be removed by a holomorphic redefinition of $S$ without loss of generality.
Here we have taken such a field basis.

In the following, we assume that a mechanism to stabilize $X$ is operative
with its vacuum expectation value fixed at $10^{10-12}$ GeV as required
by cosmological and astrophysical observations.
As suppressed by $M_{Pl}$, the interactions between $X$ and other scalar fields
rarely affect the saxion potential at $|S|\ll |X| \ll M_{Pl}$.
The Higgs potential can thus be examined by replacing $X$ with its vacuum
expectation value.
Then, the involved mass parameters are determined by the SUSY breaking
scale $M_{\rm SUSY}$ and the axion decay constant $F_a\sim |X|$.
We note that, when treating $X$ as a spurion field, the present model can be
regarded as the nMSSM,\footnote{
The nMSSM \cite{nMSSM1,nMSSM2} assumes specific discrete $R$ symmetries
to ensure the absence of large tadpoles for $S$.
A general discussion on the phenomenological aspects of the nMSSM for
small $\tan\beta$ can be found in \cite{nMSSM2,nMSSM3,DM-nMSSM,PH-nMSSM}.
Some cosmological issues have also been discussed in \cite{DM-nMSSM,PH-nMSSM}.
Neglecting small mixing with the axion superfield, the Higgs and
neutralino sectors of the PQ-NMSSM have the same phenomenological
properties as the nMSSM.
However, the cosmological properties can be different depending on the
cosmological evolution of the saxion.
A continuous symmetry to restrict couplings of $S$ in the NMSSM has been
introduced in \cite{Fayet:1974pd}, where it is explicitly broken only by
a linear superpotential of $S$.
}
where the superpotential contains an effective
tadpole for $S$:
\bea
\label{eff-PQ-NMSSM}
W_{\rm eff}=\lambda S H_uH_d + m^2_0 (1+\theta^2 B_\kappa ) S,
\eea
with $B_\kappa\sim M_{\rm SUSY}$ and
\bea
m^2_0 \sim \kappa M_{\rm SUSY} \frac{F^2_a}{M_{Pl}}.
\eea
The appearance of $m^2_0$ and $B_\kappa$ terms can be understood by
promoting $\kappa$ to a function depending on SUSY breaking fields in
a hidden sector.
It is obvious that the model does not suffer from the tadpole problem
because a tadpole for $S$ requires a higher dimensional coupling of $S$
to $X$ as dictated by $U(1)_{\rm PQ}$.
For $F_a=10^{10-12}$ GeV, $M_{\rm SUSY}\sim 1$ TeV and $\kappa$ less than
order unity, the value of $m_0$ can naturally be around the weak scale.
Hence, it is natural to expect that electroweak symmetry breaking would occur
at the correct scale.
In fact, the same spirit is shared with the Kim-Nilles mechanism
\cite{Kim-Nilles} that explains the smallness of $\mu$ in extensions
of the MSSM with $U(1)_{\rm PQ}$.
Since $S$ and $H_{u,d}$ carry $U(1)_{\rm PQ}$ charges and develop
vacuum expectation values, the Higgs and neutralino sectors have small
mixing with $X$ suppressed by $F_a$.

Let us now examine the vacuum structure of the low energy effective
theory given by (\ref{eff-PQ-NMSSM}).
Including soft SUSY breaking terms, the scalar potential
of the extended Higgs sector reads
\bea
\label{Scalar-potential}
V &=& \frac{1}{8}( g^2+g^{\prime 2}) (|H_u|^2-|H_d|^2)^2
+ \frac{1}{2}g^2 |H^\dagger_u H_d|^2
\nonumber \\
&&
+\, \left| \lambda H_uH_d + m^2_0 \right|^2
+ |\lambda|^2|S|^2 (|H_u|^2 + |H_d|^2 )
\nonumber \\
&&
+\, m^2_{H_u}|H_u|^2 + m^2_{H_d}|H_d|^2 + m^2_S |S|^2
+ \left(
A_\lambda \lambda S H_uH_d
- B_\kappa m^2_0 S
+ {\rm h.c.} \right),
\eea
where $m^2_i$ is a soft scalar mass squared, 
and $A_\lambda$ is a soft $A$-parameter.
The potential contains four complex parameters, $\lambda$, $A_\lambda$,
$m^2_0$ and $B_\kappa$.
Among them, $\lambda$ and $m^2_0$ can be made real and positive
by a field redefinition of $H_{u,d}$ and $X$.
Furthermore, if $\arg(A_\lambda)=\arg(B_\kappa)$, one can rotate away
the phases of $A_\lambda$ and $B_\kappa$ by redefining $S$.
We will assume this is the case, for which CP invariance is preserved
in the Higgs sector and there is no mixing between scalar and pseudo-scalar
fields.
From the above scalar potential, it is straightforward
to get the conditions for electroweak symmetry breaking.
Similarly as in the MSSM, two of them can be written
\bea
\frac{1}{2}M^2_Z &=&
\frac{m^2_{H_d}-m^2_{H_u}\tan^2\beta}{\tan^2\beta-1}-\mu^2_{\rm eff},
\nonumber \\
\sin2\beta &=&
\frac{2b_{\rm eff}}{m^2_{H_d}+m^2_{H_u}+2\mu^2_{\rm eff}+\lambda^2 v^2},
\eea
for $\mu_{\rm eff}$ and $b_{\rm eff}$ defined by
\bea
\mu_{\rm eff} = \lambda v_S, \quad
b_{\rm eff} = \lambda(A_\lambda v_S + m^2_0),
\eea
where $\langle |H^0_u| \rangle =v\sin\beta$ and
$\langle |H^0_d| \rangle =v\cos\beta$ with $v=174$ GeV.
The value of $|S|$ at the vacuum is fixed as
\bea
\label{v-S}
v_S =
\frac{A_\lambda \lambda v^2 \sin2\beta+2 B_\kappa m^2_0}
{2(m^2_S + \lambda^2 v^2)}.
\eea
The tree-level mass matrices for the scalar fields are presented in
the appendix \ref{mass-matrix}.

To explore the global structure of the potential, one can substitute
$S$ by the solution of $\partial_S V=0$.
Then, the Higgs potential (\ref{Scalar-potential}) is written
\bea
V = V|_{S=0} -\frac{|A_\lambda\lambda H_uH_d-B_\kappa m^2_0|^2}
{m^2_S+|\lambda|^2(|H_u|^2+|H_d|^2)},
\eea
which increases monotonically along the $D$-flat direction
$|H^0_u|=|H^0_d|$ when
\bea
\label{stability}
R_1 \geq 1\,\,{\rm and}\,\, 3R_1\geq 2+R_2,
\quad{\rm or}\quad
1\geq R^3_1 \geq R_2,
\eea
where  $R_{1,2}$ are defined by
$R_1 m^2_S =|(2\mu_{\rm eff}-A_\lambda)m^2_S
+ (2\mu_{\rm eff}-A_\lambda\sin2\beta)\lambda^2 v^2|^{2/3}$
and $R_2 m^2_S= (2\mu_{\rm eff}-A_\lambda)^2
- 2(m^2_{H_u}+m^2_{H_d}+2\mu^2_{\rm eff} -2b_{\rm eff})$.
If the above condition is not satisfied, the potential may develop another
minimum away from the weak scale.\footnote{
Actually a minimum of the potential does not lie in the $D$-flat direction
unless $m^2_{H_u}=m^2_{H_d}$.
However, as will be shown in the appendix \ref{Global}, a minimum other than
the electroweak vacuum, if exists, is located near the $D$-flat direction for
much of the parameter space.
This justifies our approach of examining the $D$-flat direction to see
when there can be another minimum.
See also \cite{Blum:2009na}, where the stability of the electroweak vacuum has
been examined within the framework of the effective Lagrangian beyond the
MSSM.
}
The involved soft parameters are then constrained by the requirement
that the electroweak vacuum should be a global minimum.
For $m^2_S \sim M^2_{\rm SUSY} \gg \mu^2_{\rm eff}$, which is the case
we shall focus on, the stability condition (\ref{stability}) requires
$A^2_\lambda \lesssim m^2_S$ or $m^2_S\lesssim A^2_\lambda\lesssim m^2_A$
with $m_A$ being the mass of the CP-odd neutral Higgs boson.
Keeping this in mind, we will consider also the case with $A_\lambda \sim M_{\rm SUSY}$,
which is favored to avoid large mixing of the SM-like Higgs scalar with
the singlet scalar when $\mu_{\rm eff}$ and $\tan\beta$ are large.

An important consequence of $U(1)_{\rm PQ}$ is the appearance
of a relatively light neutralino with a large singlino component.
This is because the PQ symmetry prevents the singlino $\tilde S$
from having a supersymmetric mass.
The lightest neutralino is mostly singlino if the masses of the
bino $\tilde B$ and wino $\tilde W$ are larger than $\lambda v$,
as is the case for $\lambda\lesssim 1$.
The neutralino mass matrix for
$(\tilde B,\tilde W^0,\tilde H^0_d,\tilde H^0_u,\tilde S)$ is given by
\bea
\hspace{-0.3cm}
{\small
\left(%
\begin{array}{ccccc}
  M_{\tilde B} & 0 & -M_Z\sin\theta_W \cos\beta
  & M_Z\sin\theta_W \sin\beta & 0 \\
  0 & M_{\tilde W} & M_Z\cos\theta_W \cos\beta
  & -M_Z\cos\theta_W \sin\beta & 0 \\
  -M_Z\sin\theta_W \cos\beta  & M_Z\cos\theta_W \cos\beta & 0
  & -\mu_{\rm eff} & -\lambda v\sin\beta  \\
  M_Z\sin\theta_W \sin\beta & -M_Z\cos\theta_W \sin\beta & -\mu_{\rm eff}
  & 0 & -\lambda v \cos\beta \\
  0 & 0 & -\lambda v \sin\beta & -\lambda v \cos\beta & 0 \\
\end{array}%
\right),
}
\eea
where $\theta_W$ is the weak mixing angle.
If we write the lightest neutralino as a linear combination of
$(\tilde B,\tilde W^0,\tilde H^0_d,\tilde H^0_u,\tilde S)$:
\bea
\tilde \chi^0_1 =
N^{\tilde B}_1 \tilde B + N^{\tilde W}_1 \tilde W^0
+ N^{\tilde H^0_d}_1 \tilde H^0_d
+ N^{\tilde H^0_u}_1 \tilde H^0_u
+ N^{\tilde S}_1 \tilde S,
\eea
then we find
\bea
&& N^{\tilde S}_1 = 1 -\frac{1}{2}(1+\cdots) \epsilon^2_{\tilde H},
\nonumber \\
&& N^{\tilde H^0_u}_1 = -\epsilon_{\tilde H}(\sin\beta + \cdots), \quad
N^{\tilde H^0_d}_1 = -\epsilon_{\tilde H}(\cos\beta + \cdots),
\nonumber \\
&&
N^{\tilde B}_1 = -\epsilon_{\tilde B}\epsilon_{\tilde H}(1+\cdots), \quad
N^{\tilde W}_1 = \epsilon_{\tilde W}\epsilon_{\tilde H}(1+\cdots),
\eea
for $\epsilon^2_{\tilde H} \ll 1$ and $|\epsilon_{\tilde B,\tilde W}|\ll 1$.
Here the epsilon parameters are defined by
\bea
\epsilon_{\tilde H} \equiv \frac{\lambda v}{\mu_{\rm eff}},
\quad
\epsilon_{\tilde B} \equiv \frac{g^\prime v\cos2\beta}{\sqrt2 M_{\tilde B}},
\quad
\epsilon_{\tilde W} \equiv \frac{g v\cos2\beta}{\sqrt2 M_{\tilde W}},
\eea
and the ellipsis indicates terms of higher orders in $\epsilon^2_{\tilde H}$
or $\epsilon_{\tilde B,\tilde W}$.
One can see that $\epsilon_{\tilde H}\neq 0$ is needed to make
$\tilde \chi^0_1$ massive through mixing.
In the following discussion, we will neglect small gaugino
components of $\tilde \chi^0_1$ since it does not change our
results substantially.
Then, one can find
\bea
\label{chi-mass}
m_{\tilde \chi^0_1} &\simeq&
2\left(\mu_{\rm eff}N^{\tilde H^0_u}_1 N^{\tilde H^0_d}_1
+ \lambda v N^{\tilde H^0_u}_1 N^{\tilde S}_1\cos\beta
+ \lambda v N^{\tilde H^0_d}_1 N^{\tilde S}_1\sin\beta \right)
\nonumber \\
&=&
\frac{\lambda^2 v^2\sin2\beta}{\mu_{\rm eff}}
\left(1 - \frac{\lambda^2v^2}{\mu^2_{\rm eff}}
+ {\cal O}\left(\frac{\lambda^4v^4}{\mu^4_{\rm eff}}\right)
\right).
\eea
As we will discuss later, the singlino-like neutralino with a small mass
can considerably change the phenomenological properties of the model.

Meanwhile, there can exist PQ messengers $\Psi+\bar \Psi$ which are
vector-like under the SM gauge group and obtain heavy masses from the
coupling $X\Psi\bar\Psi$ in the superpotential.
Such interaction can play an important role in the saxion stabilization
because it induces a radiative potential for the saxion after SUSY
breaking.
The presence of PQ messengers also helps to avoid the domain wall problem.
Let us consider $N_\Psi$ pairs of $\Psi+\bar \Psi$ forming
${\bf 5}+{\bf \bar 5}$ representation under SU(5), for which the gauge coupling
unification is preserved.
Then, the domain wall number is given by
\bea
N_{\rm DW} = |N_\Psi-6|.
\eea
This implies that the domain wall problem can be resolved for $N_\Psi=5,7$.
For other cases with $N_{\rm DW}\neq 1$, the formation of dangerous domain
walls can still be avoided if the saxion is displaced far from the origin
after the inflation ends so that the PQ symmetry is not restored at high
temperatures \cite{Soln-Domain-wall}.

\subsection{Constraints on the model parameters}

Since $S$ modifies the Higgs and neutralino sectors, it is of importance
to explore constraints on the singlet couplings $\lambda$, $A_\lambda$,
$B_\kappa$ and $m^2_S$.
Here we focus on the case with $0.1\lesssim \lambda \lesssim 1$ at
the weak scale as would be natural because an effective $\mu$ term
is generated as $\mu_{\rm eff}=\lambda S$ with $S$ fixed around $M_{\rm SUSY}$.
Let us first examine the mixing of the singlet scalar with the Higgs doublets.
After taking the rotation of $(H^0_u,H^0_d)$ by an angle $\beta$, the mass
matrix for the CP-even scalar fields has
\bea
(M^2_H)_{11} &=& M^2_Z \cos^2 2\beta + \lambda^2 v^2 \sin^2 2\beta,
\nonumber \\
(M^2_H)_{13} &=& \lambda v ( 2\mu_{\rm eff} -A_\lambda \sin2\beta),
\nonumber \\
(M^2_H)_{33} &=& m^2_S + \lambda^2 v^2,
\eea
where $(M^2_H)_{11}$ constitutes an upper bound on the mass of
the lightest CP-even Higgs boson at the tree-level.
In the following, we would like to consider the situation that the SM-like
Higgs boson has negligible contamination from the singlet scalar.
For $0.1\lesssim \lambda \lesssim 1$, this is achieved when
\bea
\frac{\mu^2_{\rm eff} }{ m^2_S}
\left|1-\frac{A_\lambda \sin2\beta}{2\mu_{\rm eff}}\right|
\ll 1,
\eea
with $m^2_S$ being of the order of $M^2_{\rm SUSY}$.
It is thus found that, if $A_\lambda$ is larger than $2\mu_{\rm eff}$,
the mixing can get a sizable suppression at some region of $\tan\beta$.
One would otherwise need $\mu^2_{\rm eff}\ll m^2_S$ to suppress
the mixing.

In the PQ-NMSSM, a stringent constraint on $\lambda$ comes from the
experimental bound on the $Z$-boson invisible decay rate because
the PQ symmetry makes $\tilde\chi^0_1$ light.
The singlino mixes with neutral Higgsinos to induce the interaction
$\bar{\tilde \chi}^0_1\sigma^\mu \tilde \chi^0_1 Z_\mu$ \cite{SUSY},
through which $Z$ can invisibly decay into pairs of the lightest
neutralino.
The coupling for this interaction is given by
\bea
g_{Z \tilde \chi^0_1 \tilde \chi^0_1}
=
\frac{g}{2\cos\theta_W}
\left(|N^{\tilde H^0_d}_1|^2-|N^{\tilde H^0_u}_1|^2 \right)
\approx \frac{g}{2\cos\theta_W}
\frac{\lambda^2 v^2\cos2\beta}{\mu^2_{\rm eff}},
\eea
where the last approximation is valid for small
$\lambda v/\mu_{\rm eff}$.
Hence, at large $\tan\beta$, the interaction gets strong
while the mass of $\tilde \chi^0_1$ becomes small.
The above coupling mediates the $Z$ decay into
$\tilde \chi^0_1$ with
\bea
\Gamma_{Z\to \tilde \chi^0_1 \tilde \chi^0_1} =
\frac{g^2_{Z\tilde \chi^0_1 \tilde \chi^0_1}}{24\pi}M_Z {\beta_Z}^3
\simeq
25 {\beta_Z}^3  \left(\frac{\lambda}{0.8}\right)^4
\left(\frac{\cos2\beta}{0.8}\right)^2
\left(\frac{200{\rm GeV}}{\mu_{\rm eff}}\right)^4
{\rm MeV},
\eea
if $M_Z>2m_{\tilde \chi^0_1}$.
Here $\beta_Z=(1-4m^2_{\tilde\chi^0_1}/M^2_Z)^{1/2}$ is the velocity
of $\tilde \chi^0_1$ in the rest frame of $Z$.
The process $Z\to \tilde \chi^0_1 \tilde \chi^0_1$ contributes
to the invisible $Z$ decay and is tightly constrained by
the LEP data to occur with a small rate,
$\Gamma_{Z\to \tilde \chi^0_1 \tilde \chi^0_1}\lesssim 2$ MeV \cite{PDG}.
This translates into
\bea
\label{Z-boson-invisible-decay}
\lambda \lesssim
0.4
\left(\frac{\mu_{\rm eff}}{200{\rm GeV}}\right)
\left(\frac{0.8}{|\cos2\beta|}\right)^{1/2}.
\eea
The above constraint on $\lambda$ around the weak scale becomes
important for large values of $\tan\beta$.
To kinematically forbid the mode $Z\to \tilde \chi^0_1 \tilde \chi^0_1$,
we need
\bea
\lambda \gtrsim 0.7
\left(\frac{\mu_{\rm eff}}{200{\rm GeV}}\right)^{1/2}
\left(\frac{0.6}{\sin2\beta}\right)^{1/2},
\eea
which is possible for $\lambda\lesssim 1$ at low $\tan\beta$.
For instance, at small $\tan\beta$ around 2, the LEP limits on the $Z$
invisible width exclude values of $\mu_{\rm eff}$ in the range between 196 GeV
and 260 GeV for $\lambda\approx0.6$ \cite{nMSSM2}.

On the other hand, for the theory to remain perturbative up to
$M_{\rm GUT}$, $\lambda$ should be small enough at the weak scale.
NMSSM models with $\lambda SH_uH_d$ usually require $\lambda$ less than
$0.7-0.8$ for $\tan\beta\gtrsim 2$.
In models with a superpotential term $S^3$, the upper bound on $\lambda$
decreases as the coupling for $S^3$ increases.
However, the situation in the PQ-NMSSM is different because $S^3$
is absent and the PQ messengers with mass $M_\Psi\propto |X|$ affect
the running of gauge couplings.
Above the scale $M_\Psi$, gauge couplings have larger values than in the MSSM
and slow down the running of Yukawa couplings.
This results in the increase of the perturbativity bound on $\lambda$ by
$\delta \lambda\lesssim 0.1$ \cite{Bound-lambda}.
A large number of light PQ messengers are favored by raising the bound,
but disfavored by the requirement of the perturbativity of gauge couplings
up to $M_{\rm GUT}$.
If there exists an extra gauge interaction, the perturbation theory would
be valid below $M_{\rm GUT}$ for a larger value of $\lambda$ at the weak
scale.

Though we do not discuss it here in detail, there is also a constraint
placed by cosmology.
If $\tilde \chi^0_1$ is the lightest sparticle, its relic abundance should
not exceed the measured amount of the dark matter.
In the present model, the production of dark matter relies on the cosmological
evolution of the saxion, which has a very flat potential generated after
SUSY breaking and thus can play some non-trivial role in cosmology.
On the other hand, the gravitino or axino can be lighter than $\tilde \chi^0_1$
depending on the mediation mechanism of SUSY breaking and on how the
saxion is stabilized.

\section{Low energy Higgs sector}

In this section, we study the low energy Higgs sector.
To see the impact of the PQ-NMSSM specific Higgs properties, we
consider the decoupling limit of the MSSM where all heavy
Higgs states decouple below $M_{\rm SUSY}$ and thus one combination
of $H_{u,d}$ behaves exactly like the SM Higgs scalar $H$.
In such a situation, we include the singlet $S$ and construct
a low energy effective theory below $M_{\rm SUSY}$ to examine
how much the model departs from the MSSM.
The modification is mainly due to $(i)$ the extra contribution
to the Higgs quartic coupling, which is a general property of
NMSSM models, and $(ii)$ the presence of a light neutralino
that is singlino-like, which is a consequence of the PQ symmetry.

\subsection{Effective theory below the SUSY breaking scale}

For $0.1\lesssim\lambda\lesssim 1$ at the weak scale,
$\mu^2_{\rm eff}\ll m^2_S$ is favored to suppress the mixing between
$H$ and the singlet scalar.
Here we consider such a case and assume that the MSSM sparticles other
than Higgsinos obtain masses of the order of $M_{\rm SUSY}$.
The singlet scalar is also assumed to have $m^2_S\sim M^2_{\rm SUSY}$.
For $\mu_{\rm eff}$ less than $M_{\rm SUSY}$, the low energy effective
theory below $M_{\rm SUSY}$ contains $\tilde H_{u,d}$ and $\tilde S$
in addition to the ordinary SM particles.
The Lagrangian relevant to our analysis is given by\footnote{
To obtain the couplings of the SM-like Higgs boson more precisely,
one needs to
know the mixing between the SM-like Higgs boson and singlet scalar.
To this end, one can replace the scalar part of (\ref{eff-L}) by
\bea
-{\cal L}_{\rm eff}|_{\rm scalar} &=&
\frac{\lambda_H-\delta \lambda_H|_{\rm tree}}{2}(|H|^2-v^2)^2
+ (m^2_S+\lambda^2 v^2)|S-v_S|^2
\nonumber \\
&&
- \left\{ \frac{\lambda(2\mu_{\rm eff}- A_\lambda\sin2\beta)}{2}
(|H|^2-v^2)(S-v_S) + {\rm c.c.}\right\},
\nonumber
\eea
and $\mu_{\rm eff}$ by $\lambda S$.
The small mixing with the singlet scalar reduces the couplings of
the Higgs boson $h$.
The reduced couplings can be obtained by taking the replacement
\bea
h \to \left(1-\frac{\lambda v|2\mu_{\rm eff}-A_\lambda \sin2\beta|}{m^2_S}
\right)h.
\nonumber
\eea
}
\bea
\hspace{-0.5cm}
\label{eff-L}
-{\cal L}_{\rm eff} = \frac{\lambda_H}{2}(|H|^2- v^2)^2
+ (y_t \bar t_R Q_L H^c + \mu_{\rm eff} \tilde H_u \tilde H_d
+ y^\prime_u H\tilde H_u \tilde S
+ y^\prime_d H^c \tilde H_d \tilde S
+ {\rm h.c.}),
\eea
where $H^c=-i\sigma_2 H^*$, and $y_t$ is the top-Yukawa coupling.
The singlino Yukawa couplings at the SUSY breaking scale are
\bea
y^\prime_u(M_{\rm SUSY}) = \lambda \cos\beta, \quad
y^\prime_d(M_{\rm SUSY}) = \lambda \sin\beta,
\eea
while the Higgs quartic coupling is given by
\bea
\lambda_H(M_{\rm SUSY})
= \frac{g^2+g^{\prime 2}}{4}\cos^2 2\beta
+ \frac{\lambda^2}{2}\sin^2 2\beta
+ \delta \lambda_H|_{\rm tree}
+ \delta \lambda_H|_{\rm loop},
\eea
where $\delta \lambda_H|_{\rm tree}$ is the threshold correction coming from
tree-level exchange of the singlet scalar,
and $\delta \lambda_H|_{\rm loop}$ is from the loops involving the stops:
\bea
\delta \lambda_H|_{\rm tree} &\simeq&
-\frac{\lambda^2(2\mu_{\rm eff}-A_\lambda\sin 2\beta)^2}{2m^2_S},
\\
\delta \lambda_H|_{\rm loop} &\simeq&
\frac{3y^4_t}{8\pi^2}\left(X_t
-\frac{X^2_t}{12} \right),
\eea
where $X_t=(A_t-\mu_{\rm eff}\cot\beta)^2/M^2_{\rm SUSY}$ with $A_t$ being
the $A$-parameter for $H_u\tilde t_R \tilde Q_L$.

The physical mass of the CP-even neutral Higgs boson $h$ can be obtained
using the relation $m^2_h=2\lambda_H v^2$.
For this, we need $\lambda_H$ renormalized at the weak scale.
In the effective theory, a low energy value of $\lambda_H$ is determined
by the renormalization group (RG) running equation:
\bea
\mu\frac{d\lambda_H}{d\mu} = \frac{1}{16\pi^2}\left(
12\lambda^2_H + 4(3y^2_t + y^{\prime 2}_u + y^{\prime 2}_d
-3A)\lambda_H + 3B-12y^4_t - 4(y^{\prime 2}_u+y^{\prime 2}_d)^2
\right),
\eea
with the parameters $A$ and $B$ defined by
\bea
4A =3g^2 + g^{\prime 2}, \quad
4B = 3g^4 + 2g^2 g^{\prime 2} + g^{\prime 4}.
\eea
Here one should note that the mixing between the neutral Higgs boson and
singlet scalar would slightly modify the running equations.

To see the qualitative properties of the Higgs mass, we make an
approximation taking into account that the dominant effects on the RG
running come from the term $y^4_t$, and also from the terms
$y^{\prime 4}_{u,d}$ if $\lambda$ is not small.
The Higgs boson mass is found to be approximately given by
\bea
\label{mh}
m^2_h &\approx& M^2_Z \cos^2 2\beta
+ \frac{3m^4_t}{4\pi^2v^2}\left( \ln\left(\frac{M^2_{\rm SUSY}}{m^2_t}\right)
+ X_t-\frac{X^2_t}{12}  \right)
\nonumber \\
&&
+\,M^2_Z\frac{2 \lambda^2}{g^2+g^{\prime 2}} \left( \sin^2 2\beta
-\frac{(2\mu_{\rm eff}-A_\lambda \sin2\beta)^2}{m^2_S}
+ \frac{\lambda^2}{4\pi^2}\ln\left(\frac{M^2_{\rm SUSY}}{\mu^2_{\rm eff}}\right)
\right),
\eea
for $m_t=y_t v$ and $\mu_{\rm eff}\gtrsim m_h$.
The first line is the well-known result for the Higgs boson mass in the MSSM
\cite{Higgs-mass-MSSM1,Higgs-mass-MSSM2}.
On the other hand, those in the second line correspond to the additional
contributions arising due to $S$, i.e. as a consequence of
the extra Higgs quartic coupling $\lambda^2|H_uH_d|^2$,
the mixing between the singlet scalar and neutral Higgs boson, and
the singlino Yukawa interactions affecting the running of the
Higgs quartic coupling at low energy scales.
The last two contributions are approximately estimated as
\bea
\delta m_h|_{\rm mix} &\approx& -10\left(\frac{130{\rm GeV}}{m_h}\right)
\left(\frac{30}{m^2_S/\mu^2_{\rm eff}}\right)
\left(1-\frac{A_\lambda\sin2\beta}{2\mu_{\rm eff}}\right)^2
\left(\frac{\lambda}{0.8}\right)^2{\rm GeV},
\\
\delta m_h|_{\rm rad} &\approx& 4.1 \left(\frac{130{\rm GeV}}{m_h}\right)
\left(\frac{\ln(M^2_{\rm SUSY}/\mu^2_{\rm eff})}{\ln30}\right)
\left(\frac{\lambda}{0.8}\right)^4{\rm GeV}.
\eea
A negative contribution to $m_h$ from the mixing with the singlet scalar
is present in any NMSSM model.
In the PQ-NMSSM, $\mu^2_{\rm eff} \ll m^2_S$ leads to a large suppression
of this effect for $A_\lambda\lesssim \mu_{\rm eff}$.
For $A_\lambda\gtrsim 2\mu_{\rm eff}$, small mixing can still be obtained
at some values of $\tan\beta$.
One should also note that there is a PQ-NMSSM specific contribution
$\delta m_h|_{\rm rad}$ arising because the PQ symmetry makes
the lightest neutralino get a relatively small mass.\footnote{
See also \cite{Nakayama} for a similar discussion in a singlet extension
of the MSSM having a relatively light neutralino.
However, in our situation, the LEP bound on the invisible $Z$-boson decay width
excludes large values of $\lambda$ at large $\tan\beta$.
}
This positive contribution is insensitive to $\tan\beta$, and becomes
important for a small value of $\mu_{\rm eff}/M_{\rm SUSY}$ contrary to
$\delta m_h|_{\rm mix}$.

\begin{figure}[t]
\begin{center}
\begin{minipage}{15cm}
\centerline{
{\hspace*{0cm}\epsfig{figure=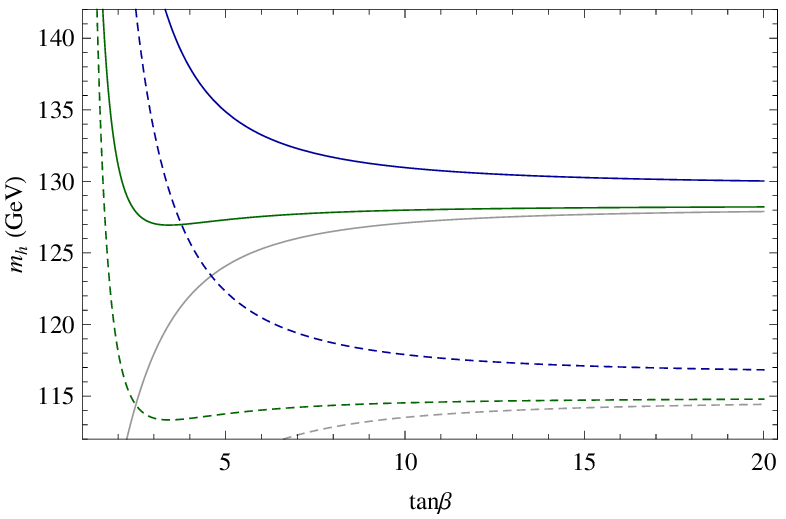,angle=0,width=7.2cm}}
{\hspace*{0.4cm}\epsfig{figure=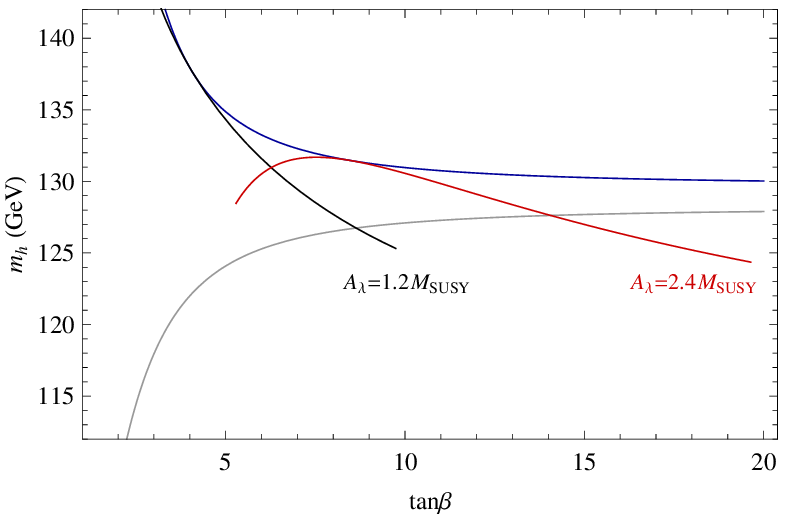,angle=0,width=7.2cm}}
}
\caption{
The mass of the SM-like Higgs boson for $M_{\rm SUSY}=1.5$ TeV
in the PQ-NMSSM.
Here $\lambda$ is taken to be the maximum value satisfying
the bound $\Gamma_{Z\to \tilde\chi^0_1\tilde\chi^0_1}\lesssim 2$ MeV.
The solid curve in the left panel is the upper bound on $m_h$
for $\mu_{\rm eff}=240$ GeV (green) and $\mu_{\rm eff}=420$ GeV (blue),
which is obtained for no mixing case $\delta m_h|_{\rm mix}=0$.
The gray line is the MSSM value of $m_h$.
In the left panel, the dashed lines are for $X_t=0$ while the solid
ones for $X_t=6$.
Meanwhile, the right panel shows $m_h$ for $X_t=6$,
$\mu_{\rm eff}=420$ GeV and a given value of $A_\lambda$:
the black curve is for $A_\lambda=1.2 M_{\rm SUSY}$ while
the red one for $A_\lambda=2.4 M_{\rm SUSY}$.
}
\label{fig:Higgs-mass}
\end{minipage}
\end{center}
\end{figure}

In Fig. \ref{fig:Higgs-mass}, we show the upper bound on $m_h$ in the
PQ-NMSSM, which is obtained taking the maximum value of
$\lambda$ allowed by the constraint (\ref{Z-boson-invisible-decay})
from the $Z$ invisible decay.
Here we have solved the RG equation to get $\lambda_H$ at the weak
scale, and restricted $\lambda$ to be less than unity as would be
necessary to maintain its perturbativity up to $M_{\rm GUT}$.
In the left panel, the value of $m_h$ is shown for $\delta m_h|_{\rm mix}=0$.
While the MSSM generates $m_h\simeq 115$ GeV (128 GeV) at
$\tan\beta\gtrsim 10$ for $X_t=0$ (6) and $M_{\rm SUSY}=1.5$ TeV,
the additional contribution from $\lambda$ can lead to $m_h$ larger than
115 GeV also at low $\tan\beta$ as in other NMSSM models.
The loops of stops involving $A_t$ can further increase $m_h$.
The maximum comes at $X_t=6$.
It is also important to note that, when $\mu_{\rm eff}\gtrsim 400$ GeV,
$m_h$ can be raised by a few GeV from the MSSM value even at large
$\tan\beta$ owing to the PQ-NMSSM specific contribution
$\delta m_h|_{\rm rad}$.
On the other hand, the right panel shows the value of $m_h$ for a given value
of $A_\lambda$.
For $A_\lambda\gtrsim 2\mu_{\rm eff}$, the mixing effect is suppressed
only at some limited region of $\tan\beta$.
In the figure, we consider values of $\tan\beta$ giving
$(2\mu_{\rm eff}-A_\lambda \sin2\beta)^2/m^2_S$ less than 0.1.

\subsection{Phenomenological aspects}

Since there appears a light neutralino as a consequence of the
PQ symmetry, the Higgs boson $h$ can invisibly decay into pairs
of the lightest neutralino.
This process is mediated by the Yukawa interaction
$h\tilde \chi^0_1\tilde \chi^0_1$, which is generated due to
the mixing between $\tilde S$ and $\tilde H^0_{u,d}$ and
has a coupling given by
\bea
y_{h\tilde \chi^0_1\tilde \chi^0_1}
=
-\sqrt2 \lambda \left( N^{\tilde H^0_d}_1 \sin\beta
+ N^{\tilde H^0_u}_1\cos\beta \right)
\approx
\frac{\sqrt2\lambda^2 v\sin2\beta}{\mu_{\rm eff}},
\eea
where the approximation is valid for small $\lambda v/\mu_{\rm eff}$.
The above coupling becomes negligible at large $\tan\beta$.
If dominates, such non-standard invisible decay would make the
Higgs discovery at hadron colliders much more difficult.

Recent LHC data have excluded the Higgs boson with SM properties
in the mass range between 141 GeV and 476 GeV at the 95\% confidence
level \cite{SM-Higgs-LHC}.
For $h$ with mass lighter than 141 GeV, the main processes for its decay
are $h\to b\bar b$ and $h\to WW^*,$ $ZZ^*$ \cite{Higgs-decay}.
The Higgs boson $h$ in the PQ-NMSSM, which would have a small singlet
component for $\mu^2_{\rm eff}/m^2_S\ll 1$, can decay through a
non-standard mode $h\to \tilde \chi^0_1 \tilde \chi^0_1$ \cite{nMSSM3}.
If it is kinematically accessible, the process
$h\to \tilde \chi^0_1\tilde \chi^0_1$ takes place with the relative decay
strength
\bea
\hspace{-0.5cm}
\frac{\Gamma_{h\to\tilde \chi^0_1\tilde \chi^0_1}}{\Gamma_{h\to b\bar b}}
&\simeq&
\frac{1}{3}\left(\frac{y_{h\tilde \chi^0_1\tilde \chi^0_1}}{m_b/v}\right)^2
\simeq
128
\left(\frac{\lambda}{0.8}\right)^4
\left(\frac{\sin2\beta}{0.6}\right)^2
\left(\frac{200{\rm GeV}}{\mu_{\rm eff}}\right)^2,
\\
\hspace{-0.5cm}
\frac{\Gamma_{h\to\tilde \chi^0_1\tilde \chi^0_1}}{\Gamma_{h\to WW^*}}
&\simeq&
\frac{32\pi^2}{3R(x)}\left(\frac{y_{h\tilde \chi^0_1\tilde \chi^0_1}}{g^2}
\right)^2
\simeq
440\left(\frac{0.3}{R(x)}\right)
\left(\frac{\lambda}{0.8}\right)^4
\left(\frac{\sin2\beta}{0.6}\right)^2
\left(\frac{200{\rm GeV}}{\mu_{\rm eff}}\right)^2,
\eea
where we have ignored the masses of the final states,
and $R(x)$ is defined by
\bea
R(x)  &=& \frac{3(1-8x+20x^2)}{(4x-1)^{1/2}}\arccos
\left( \frac{3x-1}{2x^{3/2}} \right)
\nonumber \\
&&
-\,\frac{1-x}{2x}(2-13x+47x^2) -\frac{3}{2}(1-6x+4x^2)\ln x,
\eea
with $x=M^2_W/m^2_h$.
The decay rate for the process $h\to ZZ^*$ is similar to $\Gamma_{h\to WW^*}$.
When $h$ has a sizable singlet component, the Higgs decay width
for each process is modified, but the ratio between decay
widths remains the same up to small correction arising due to that
$y_{h\tilde\chi^0_1\tilde \chi^0_1}$ receives contribution not only
from $H^0\tilde H^0_{u,d}\tilde S$ but also from
$S\tilde H^0_u\tilde H^0_d$.
The Higgs invisible decay to neutralinos would not dominate the SM decay
processes either if $\lambda$ is strong enough to make $\tilde\chi^0_1$
heavier than $m_h/2$:
\bea
\lambda \gtrsim 0.85
\left(\frac{m_h}{130{\rm GeV}}\right)^{1/2}
\left(\frac{\mu_{\rm eff}}{200{\rm GeV}}\right)^{1/2}
\left(\frac{0.6}{\sin2\beta}\right)^{1/2},
\eea
or if $\lambda$ is small enough to suppress the Yukawa coupling of
$\tilde \chi^0_1$ to the Higgs boson:
${\rm Br}(h\to\tilde \chi^0_1\tilde \chi^0_1)$ is less than 0.5 for
\bea
\lambda \lesssim 0.27
\left(\frac{\mu_{\rm eff}}{200{\rm GeV}}\right)^{1/2}
\left(\frac{0.6}{\sin2\beta}\right)^{1/2}.
\eea
Here we have naively estimated the value of $\lambda$ required for
$m_{\tilde \chi^0_1}>m_h/2$ by taking the leading term in (\ref{chi-mass}),
which is expanded in powers of $\lambda^2 v^2/\mu^2_{\rm eff}$.
It is interesting to see that, in a low $\tan\beta$ region, the Higgs
boson decays mainly through the invisible channel $h\to \tilde \chi^0_1\tilde \chi^0_1$
for $\lambda\gtrsim 0.4$ and $\mu_{\rm eff}\lesssim 400$ GeV.
A large $\mu_{\rm eff}$ can weaken this decay mode, but would lead to
large mixing between $H^0$ and $S$.
On the other hand, for $\tan\beta\gtrsim 10$, the constraint from the
invisible $Z$-boson decay (\ref{Z-boson-invisible-decay}) requires
$\lambda\lesssim 0.36\times(\mu_{\rm eff}/200{\rm GeV})$.
Thus, in this case, ${\rm Br}(h\to\tilde \chi^0_1\tilde \chi^0_1)$ cannot
be larger than 0.5 if $\mu_{\rm eff}$ is smaller than about 360 GeV.

\begin{figure}[t]
\begin{center}
\begin{minipage}{15cm}
\centerline{
{\hspace*{0cm}\epsfig{figure=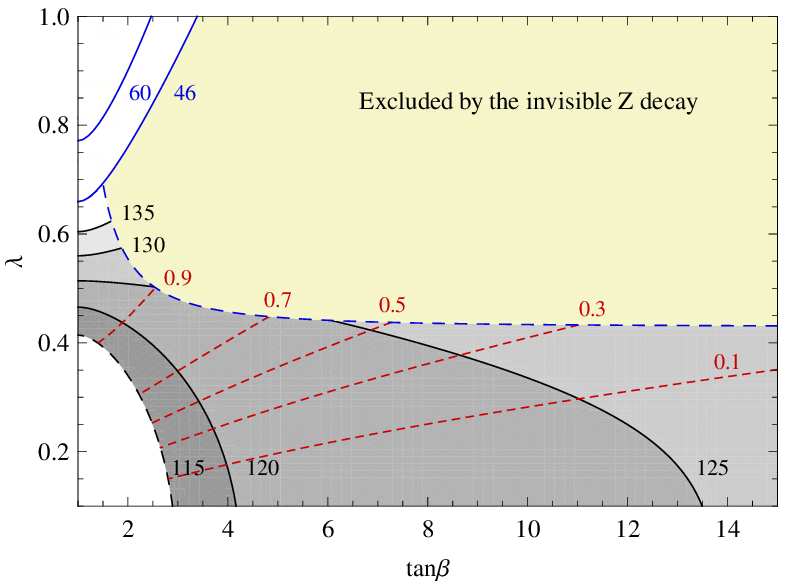,angle=0,width=7.3cm}}
{\hspace*{0.2cm}\epsfig{figure=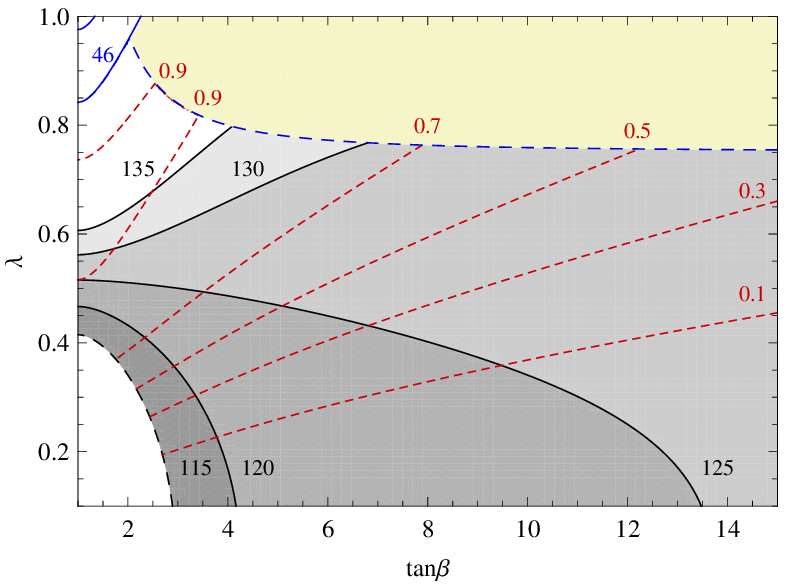,angle=0,width=7.3cm}}
}
\caption{
The branching ratio of non-standard mode
$h\to\tilde \chi^0_1\tilde \chi^0_1$
for $\mu_{\rm eff}=240$ GeV (left) and $\mu_{\rm eff}=420$ GeV (right).
The dashed red line is the contour for
${\rm Br}(h\to\tilde \chi^0_1\tilde \chi^0_1)$.
We also show a contour plot for $m_{\tilde \chi^0_1}$ larger than $M_Z/2$,
which is given in blue.
The yellow region is excluded by the experimental bound on the invisible
$Z$ decay width.
Meanwhile, the black contour shows the value of $m_h$ obtained for
$\delta m_h|_{\rm mix}=0$ in the case with $M_{\rm SUSY}=1.5$
TeV and $X_t=6$.
The Higgs boson has a mass larger than 115 GeV above the dashed black line.
In the figure, masses are given in the GeV unit.
}
\label{fig:Higgs-decay}
\end{minipage}
\end{center}
%
\begin{center}
\begin{minipage}{15cm}
\centerline{
{\hspace*{0cm}\epsfig{figure=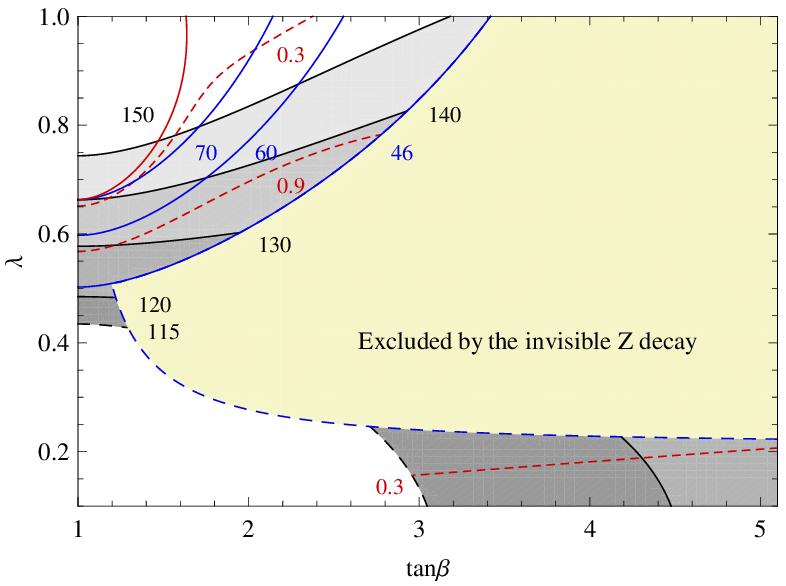,angle=0,width=7.3cm}}
{\hspace*{0.2cm}\epsfig{figure=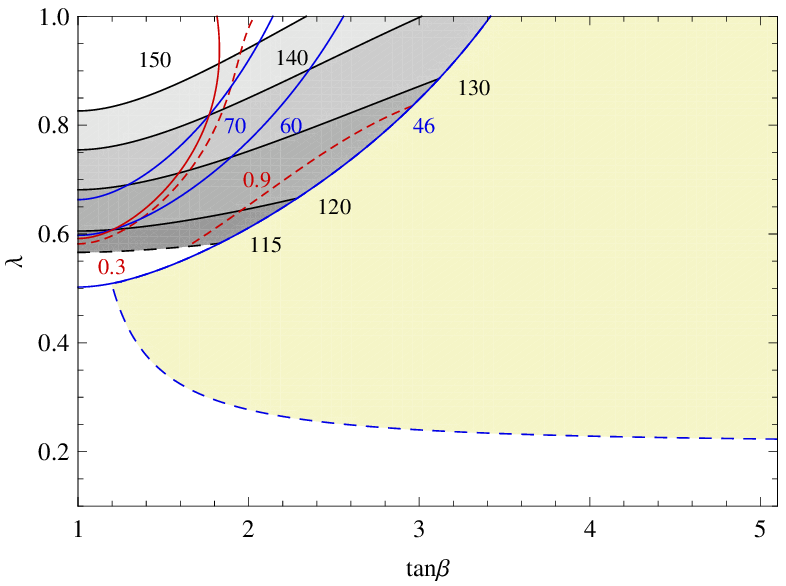,angle=0,width=7.3cm}}
}
\caption{
The parameter region where $Z\to \tilde\chi^0_1\tilde\chi^0_1$ is
kinematically forbidden.
The blue contour shows the value of $m_{\tilde \chi^0_1}$ for
$\mu_{\rm eff}=120$ GeV.
We also show the value of $m_h$ by a black contour for the case
with $M_{\rm SUSY}=1$ TeV, $X_t=6$ (left) and $X_t=0$ (right).
The masses are given in the GeV unit.
The dashed red line is a contour for ${\rm Br}(h\to \tilde \chi^0_1
\tilde \chi^0_1)$.
Above the solid red line, the lightest neutralino obtains a mass larger
than $m_h/2$.
}
\label{fig:small-mu}
\end{minipage}
\end{center}
\end{figure}

Fig. \ref{fig:Higgs-decay} shows the branching ratio for
the invisible Higgs decay $h\to \tilde\chi^0_1\tilde\chi^0_1$
in the $(\tan\beta,\lambda)$ plane.
In the figure, the yellow region is excluded by the experimental bound
$\Gamma_{Z\to\tilde\chi^0_1\tilde\chi^0_1}\lesssim 2$ MeV,
and the black contour is the value of $m_h$ obtained for
$M_{\rm SUSY}=1.5$ TeV and $X_t=6$ with $\delta m_h|_{\rm mix}=0$.
Here we have taken $\mu_{\rm eff}=240$ GeV for the left
plot, and $\mu_{\rm eff}=420$ GeV for the right plot.
Since there is an extra contribution to $m_h$ from $\lambda$
as (\ref{mh}), the Higgs mass can be raised above $115$ GeV at
low $\tan\beta$.
Notice also that the contribution $\delta m_h|_{\rm rad}$
raises $m_h$ by a few GeV even at $\tan\beta\gtrsim 10$
compared to the MSSM value that corresponds to the $\lambda=0$ case.
The process $h\to \tilde\chi^0_1\tilde\chi^0_1$ can be the
dominant mode of the Higgs decay at $\tan\beta\lesssim 10$, but is
suppressed at large $\tan\beta$.

On the other hand, for small $\mu_{\rm eff}$ less than $\lambda v$,
the lightest neutralino is a sizable mixture of $\tilde S$ and
$\tilde H^0_{u,d}$ at low $\tan\beta$.
In this case, it can acquire a mass larger than $M_Z/2$ so that
the invisible decay $Z\to \tilde\chi^0_1\tilde\chi^0_1$ is kinematically
forbidden.
For some parameter region, it is possible for $\tilde \chi^0_1$ to get
a mass even larger than $m_h/2$.
Otherwise, $h$ would decay dominantly through the invisible process
$h\to \tilde \chi^0_1\tilde \chi^0_1$ at low $\tan\beta$ because
$\tilde \chi^0_1$ has a sizable Higgsino component.
One should also note that the contribution from $\lambda$ can raise
$m_h$ well above 115 GeV for $M_{\rm SUSY}\lesssim 1$ TeV at $\tan\beta\lesssim 3$.
In Fig. \ref{fig:small-mu}, we show the region of $\lambda$ and $\tan\beta$
where $\tilde\chi^0_1$ has a mass larger than $M_Z/2$.
The Higgs boson mass is also shown for the case with $M_{\rm SUSY}=1$ TeV.

\section{UV completion}

When one considers GUT models to UV complete the PQ-NMSSM,
an important issue is how the GUT partners of the MSSM Higgs
doublets, which also carry a PQ charge, acquire heavy masses.
We point out that the PQ-NMSSM can emerge as a low energy
effective theory of a missing-partner model for supersymmetric
SU(5) GUT \cite{Missing-partner}.\footnote{
A 5 dimensional SU(5) unified theory can also yield the PQ-NMSSM
when compactified on $S^1/(Z_2\times Z^\prime_2)$ orbifold
\cite{Extra-dim,GUT-in-extra-dim}.
For instance, one can introduce a pair of Higgs hypermultiplet
${\cal H}+\bar {\cal H}$ which form $5+\bar 5$ representation of SU(5)
and carry a PQ charge $-1$.
Then, doublet-triplet splitting is achieved taking the orbifold projection
such that ${\cal H}$ ($\bar{\cal H}$) has a SU(2) doublet Higgs chiral
multiplet transforming as $(+,+)$ under $Z_2\times Z^\prime_2$ and
a triplet Higgs with $(+,-)$.
This also leads to the terms (\ref{PQ-NMSSM}) for the PQ-NMSSM below
the compactification scale.
}
The missing-partner model has been considered to explain
a large mass splitting of the SU(2) doublet and color triplet Higgses.
The idea is to introduce higher dimensional representations that contain
Higgs triplets but no doublets.
Then, if a mass term $H_5 \bar H_{\bar 5}$ for ${\bf 5}+{\bf \bar 5}$ Higgs
multiplets is absent in the superpotential, the doublet-triplet splitting
can be achieved without fine-tuning from the interactions of $H_5$ and
$\bar H_{\bar 5}$ with the higher dimensional Higgs multiplets.
This is possible because the superpotential is not renormalized in
perturbation theory.

The Higgs sector of the original model consists of the chiral multiplets,
$\Sigma({\bf 75})$, $\theta({\bf 50})+\bar\theta(\overline{\bf 50})$
and $H({\bf 5})+\bar H({\bf \bar 5})$.
To incorporate the PQ symmetry without spoiling the missing-partner mechanism,
we modify the model by introducing three pairs of ${\bf 50}+\overline{\bf 50}$
and ${\bf 5}+{\bf \bar 5}$ multiplets, and also three SU(5) singlets:
\bea
\Sigma({\bf 75}),\quad
\theta_i({\bf 50})+\bar\theta_i(\overline{\bf 50}),\quad
H_i({\bf 5})+\bar H_i({\bf \bar5}), \quad
X_i({\bf 1}),
\eea
where $i=(1,2,3)$, and $U(1)_{\rm PQ}$ charges are assigned as
\bea
\label{PQ-charge-assignment}
&&
\Sigma(0), \quad
X_1(q), \quad
X_2(-3q), \quad
X_3(2q),
\nonumber \\
&&
\theta_1(-p)+\bar\theta_1(p), \quad
\theta_2(-3q-p)+\bar\theta_2(3q+p), \quad
\theta_3(-q-p)+\bar\theta_3(q+p),
\nonumber \\
&&
H_1(p)+\bar H_1(-q-p), \quad
H_2(3q+p)+\bar H_2(-p), \quad
H_3(q+p)+\bar H_3(-3q-p).
\eea
This model seems similar to the minimal model with $U(1)_{\rm PQ}$
considered in \cite{PQ-missing-partner}, but it turns out that more than
two pairs of chiral multiplets are needed to obtain the PQ-NMSSM as
an effective theory below $M_{\rm GUT}$.
The missing-partner mechanism is implemented by the following PQ-invariant
superpotential terms
\bea
\label{missing-partner}
W = \frac{1}{2}M{\rm Tr}(\Sigma^2)
+ \frac{1}{3} a{\rm Tr}(\Sigma^3)
+ b_i \theta_i \Sigma H_i
+ c_i \bar\theta_i \Sigma \bar H_{i+1}
+ \tilde M_i \bar\theta_i \theta_i,
\eea
with the identification $\bar H_4=\bar H_1$.
As in the original model, the vacuum expectation value of $\Sigma$
breaks SU(5) to the SM gauge groups, and gives rise to the triplet mass
terms.
This becomes clear after integrating out the heavy triplets in
$\theta_i+\bar\theta_i$, which leads to
\bea
W_{\rm eff} = M^c_i H^c_i\bar H^c_{i+1},
\eea
where $M^c_i\sim M^2_{\rm GUT}/\tilde M_i$,
and $H^c_i$ denotes the color-triplet from $H_i$.
Because the ${\bf 50}$ representation does not contain Higgs doublets,
no doublet mass terms are generated from the superpotential
(\ref{missing-partner}), and the three pairs of doublet Higgses remain
massless.
Mass terms for these doublet Higgses arise from
\bea
\label{doublet-masses}
W =\lambda_i X_i H_i \bar H_i + \frac{\xi}{M_{Pl}}X^3_1 X_2,
\eea
where we have chosen a basis of $X_3$ such that $X^2_1 H_3\bar H_3$
is removed in the superpotential.
Including soft SUSY breaking terms for the gauge singlet scalars
\bea
\label{missing-PQ-scale}
-{\cal L}_{\rm soft} = m^2_{X_i} |X_i|^2
+ \left(A_\xi \frac{\xi}{M_{Pl}}X^3_1 X_2
+{\rm h.c.}\right),
\eea
the second term in the above superpotential fixes $X_{1,2}$ at
\bea
|X_{1,2}|^2 \approx \frac{M_{Pl}(-m^2_{X_1})^{1/2}}{\xi}
\sim \frac{M_{Pl}M_{\rm SUSY}}{\xi},
\eea
thereby leading to that $U(1)_{\rm PQ}$ is spontaneously broken
around $10^{11}$ GeV for $\xi \sim 1$, and there appears
the axion which is a mixture of $\arg(X_{1,2})$.
Here we have assumed $m^2_{X_1}<0$.
Hence, the doublet Higgses in $H_{1,2}+\bar H_{1,2}$ obtain
large masses from the vacuum expectation value of $X_{1,2}$, respectively.
On the other hand, the doublet Higgses in $H_3+\bar H_3$ remain
massless until $X_3$ acquires a nonzero vacuum expectation value.

Notice that the PQ charge assignment (\ref{PQ-charge-assignment})
allows direct mass terms $H_i\bar H_{i+1}$ and the Yukawa term
$X_1X_2X_3$ in the renormalizable superpotential.
These terms should be absent in order for the missing-partner
mechanism to work and for only one pair of Higgs doublets to remain
light.
Once we do not put these superpotential terms, radiative corrections
will not change the situation owing to supersymmetry.\footnote{
One can assign a different PQ charge to $X_2$ and the Higgs multiplets
to forbid a superpotential term $X_1X_2X_3$.
Then, other mechanism is needed to fix the PQ breaking scale because
the term $X^3_1X_2$ in the scalar potential (\ref{missing-PQ-scale})
is not allowed.
}
It is also important to note that the model possesses two global
$U(1)$ symmetries associated with the independent charges $p$ and $q$.
To eliminate one of them, as was considered in \cite{PQ-missing-partner},
we introduce three right-handed neutrino multiplets $N({\bf 1})$
that implement the conventional see-saw mechanism through the
superpotential terms $NLH_u +X_i NN$ with $i=1$ or 2.
The PQ charges are then fixed as $5p=-8q$ when the Majorana masses
for $N$ arise from $X_1NN$, and $5p=-12q$ if one instead chooses
$X_2NN$.

It now becomes apparent that the missing-partner model with the superpotential
terms (\ref{missing-partner}) and (\ref{doublet-masses}) leads to
the PQ-NMSSM.
The doublet Higgses in $H_3+\bar H_3$ correspond to the ordinary MSSM
Higgses, while $X_{1,2}$ and $X_3$ play the role of $X$ and $S$,
respectively.
In the model, the higher dimensional operators
\bea
{\cal L} = \int d^4\theta\left(
\kappa_1 \frac{X^{*2}_1X_3}{M_{Pl}}
+ \kappa_2 \frac{X_1X_2X_3}{M_{Pl}} \right) + {\rm h.c.}
\eea
can generate an effective tadpole term for $X_3$ as
\bea
W_{\rm eff} = \tilde m^2_0 X_3,
\eea
where $\tilde m^2_0 \sim M^2_{\rm SUSY}$ for $\kappa_{1,2}\sim 1$ and
$\xi\sim 1$.
Then, $X_3$ is naturally expected to get a vacuum expectation value around the
weak scale for $M_{\rm SUSY}\sim 1$ TeV.
However, if PQ-breaking mass terms $\theta_i\bar\theta_j$
with $i\neq j$ are present, the loops of heavy triplets would generate
large tadpoles for $X_3$.
This implies that $U(1)_{\rm PQ}$ is crucial to avoid the tadpole problem.
The PQ symmetry plays an important role also in suppressing dangerous higher
dimensional operators leading to too rapid proton decays.
The triplet Higgses mediate dimension 5 operators violating the baryon
number \cite{proton-decay}, which carry a nonzero PQ charge and therefore
are further suppressed by a small factor
$X_1X_2/(M^c_1 M^c_2)\sim M_{\rm SUSY}M_{Pl}/(M^c_1 M^c_2)$ compared to
those in the minimal SU(5) GUT model.

Let us finally discuss the difference from the model of
\cite{PQ-missing-partner}.
That model contains two pairs of Higgs doublets $H_f+\bar H_f$ and
$H^\prime_f +\bar H^\prime_f$ which are vector-like also under $U(1)_{\rm PQ}$.
One pair of them becomes heavy through $PH_f\bar H^\prime_f$ for $P$ being
a $U(1)_{\rm PQ}$ breaking gauge singlet field, and the other remains light.
If one introduces an additional singlet $P^\prime$ having a term
$P^\prime H^\prime_f \bar H_f$ in the superpotential, $P^\prime$ necessarily
carries a PQ charge such that $PP^\prime$ is invariant under $U(1)_{\rm PQ}$
transformations.
Thus, even if one omits $PP^\prime$ in the superpotential,
a K\"ahler potential term $PP^\prime$ would induce too large tadpole
term for $P^\prime$.
This makes it difficult for $P^\prime$ to play the role of $S$ in
the PQ-NMSSM.

\section{Conclusions}

Extended to incorporate the PQ mechanism solving the strong CP problem,
the NMSSM becomes compatible with the grand unification since the PQ
symmetry forbids large tadpoles for the SM singlet $S$ to be generated
from loops of heavy fields coupling to $S$.
Another important property of the PQ-NMSSM is that all the mass
parameters are determined by the SUSY breaking scale $M_{\rm SUSY}$
and $F^2_a/M_{Pl}$ with $F_a$ being the axion decay constant.
Thus, the electroweak symmetry breaking is naturally achieved at
the correct scale.
Furthermore, the model can avoid the domain wall problem in the presence
of the PQ messengers.

An important consequence of the PQ symmetry is that the lightest
neutralino is singlino-like with a small Higgsino admixture, and
is relatively light compared to other sparticles.
The Higgsino component is determined by the coupling of $S$ to the Higgs
doublets, which is constrained by the LEP bound on the invisible $Z$-boson
decay width.
This constraint becomes severe at large $\tan\beta$.
Meanwhile, the SM-like Higgs boson decays mainly through the conventional decay
modes at large $\tan\beta$ and in a portion of parameter space for small
values of $\tan\beta$.
The decay of the Higgs boson into a pair of the lightest neutralino
can be the main mode at low $\tan\beta$, for which case the Higgs search
at colliders will be modified.
Also important is that the SM-like Higgs mass receives an additional
positive contribution from the loops involving the singlino Yukawa coupling.
This PQ-NMSSM specific contribution can lead to a significant increase of the
Higgs boson mass by a few GeV even at large $\tan\beta$ compared to the MSSM.

We found that the PQ-NMSSM is realized as a low energy effective
theory of a missing-partner model for supersymmetric SU(5) GUT
with the PQ symmetry, which solves the doublet-triplet splitting problem
and the proton decay problem.
It is interesting to note that such a UV completion achieves
the relation $F_a\sim \sqrt{M_{\rm SUSY}M_{Pl}}$.
Hence, all the mass parameters of the resulting PQ-NMSSM
have values of the order of $M_{\rm SUSY}$.

\vskip 0.5cm
{\bf Note added}

After submitting the manuscript, the ATLAS and CMS collaborations
at the LHC reported their updated results in the Higgs search \cite{Higgs-LHC},
which may indicate a SM-like Higgs boson with mass around 125 GeV.
To explain a 125 GeV Higgs mass within the MSSM, we need large stop mixing
or heavy stops with mass larger than about 10 TeV.
The PQ-NMSSM improves the situation because the Higgs mass receives an
additional positive contribution, which can be of a few GeV even at
large $\tan\beta$.

\acknowledgments

This work is supported by Grants-in-Aid for Scientific Research from
the Ministry of Education, Science, Sports, and Culture (MEXT),
Japan, No. 23104008 and No. 23540283,
and in part by the JSPS Grant-in-Aid 21-09224 (K.S.J), and also by
Tohoku University International Advanced Research and Education
Organization (Y.S.).

\appendix

\section{\label{mass-matrix}Mass matrix}

In this appendix, we present the tree-level mass matrices
for the neutral scalar fields.
After rotating the upper left $2\times2$ submatrix of
the mass matrix for the CP even scalars, one obtains
\bea
(M^2_H)_{11} &=& M^2_Z \cos^2 2\beta +\lambda^2 v^2 \sin^2 2\beta,
\nonumber \\
(M^2_H)_{22} &=& \frac{2b_{\rm eff}}{\sin 2\beta}
+ (M^2_Z-\lambda^2 v^2)\sin^2 2\beta,
\nonumber \\
(M^2_H)_{33} &=& m^2_S + \lambda^2 v^2,
\nonumber \\
(M^2_H)_{12,21} &=& \frac{1}{2} (M^2_Z-\lambda^2 v^2) \sin4\beta,
\nonumber \\
(M^2_H)_{13,31} &=& \lambda v (2\mu_{\rm eff}-A_\lambda\sin2\beta),
\nonumber \\
(M^2_H)_{23,32} &=& \lambda v A_\lambda \cos2\beta.
\eea
The mass matrix for the pseudoscalar fields is given by
\bea
M^2_A =\left(%
\begin{array}{ccc}
  b_{\rm eff}\cot\beta & b_{\rm eff} & \lambda v A_\lambda \cos\beta \\
  b_{\rm eff} & b_{\rm eff}\tan\beta & \lambda v A_\lambda \sin\beta \\
  \lambda v A_\lambda \cos\beta & \lambda v A_\lambda \sin\beta
  & m^2_S + \lambda^2 v^2 \\
\end{array}%
\right).
\eea
It is easy to see that there are one massless mode, which
is absorbed into gauge boson, and two massive CP odd scalars:
\bea
M^2_{A_{1,2}} =
\frac{b_{\rm eff}}{\sin2\beta}
+ \frac{1}{2}(m^2_S+\lambda^2 v^2)
\pm
\sqrt{
\left(
\frac{b_{\rm eff}}{\sin2\beta}
- \frac{1}{2}(m^2_S+\lambda^2 v^2)
\right)^2
+ A^2_\lambda \lambda^2 v^2}.
\eea
Using the stationary condition (\ref{v-S}), one can find that
$M_{A_1}=0$ if $\kappa=0$.

\section{\label{Global}Global structure of the Higgs potential}

In the PQ-NMSSM, where the Higgs sector is extended to include the singlet $S$,
the Higgs potential may develop another minimum away from the weak scale.
The model parameters are constrained to avoid such a minimum since
it would generally appear at a field value similar to or larger than $M_{\rm SUSY}$
and thus be deeper than the electroweak vacuum.
For the case with $m^2_S>0$ and $\lambda\lesssim 1$, we shall show that
the region $\lambda^2(|H^0_u|^2+|H^0_d|^2)\sim M^2_{\rm SUSY}$ somewhat
near the $D$-flat direction is potentially dangerous for much of the parameter
space.
This argues that the condition to avoid a deeper minimum can approximately be examined
by looking at the shape of the potential along the $D$-flat direction.

After integrating out $S$ by the minimization condition, the Higgs potential
reads
\bea
V &=& \left( \frac{g^2+g^{\prime 2}}{2}\cos^22\theta
+ \lambda^2 \sin^2 2\theta \right)\phi^4
- \frac{(A_\lambda \lambda \phi^2 \sin2\theta+B_\kappa m^2_0)^2}{2\lambda^2\phi^2+m^2_S}
\nonumber \\
&&
+\, 2\left(m^2_{H_u}\sin^2\theta+m^2_{H_d}\cos^2\theta
-\lambda m^2_0 \sin2\theta \right)\phi^2
+ {\rm constant},
\eea
where $|H^0_u| = \sqrt2 \phi \sin\theta$ and
$|H^0_d| = \sqrt2 \phi \cos\theta$ with $0\leq \theta <\pi/2$ and $0\leq \phi$.
At very large values of $\phi$, the first term becomes dominant and lifts the potential
along the $\phi$-direction.
It is also straightforward to see that $\partial_\phi V=0$ when $\phi=0$ or when
\bea
\left(\sin^22\theta + \frac{g^2+g^{\prime 2}}{2\lambda^2}\cos^22\theta\right)
\left( 2\lambda^2\phi^2+m^2_S \right)^3
- k m^2_S \left( 2\lambda^2\phi^2+m^2_S \right)^2
&&
\nonumber \\
+\, 2\left(
\frac{A_\lambda m^2_S}{2}\sin2\theta - \lambda B_\kappa m^2_0 \right)^2
&=& 0,
\eea
where $k$ is a function of $\theta$,
\bea
\hspace{-0.5cm}
k = \left(1+\frac{A^2_\lambda}{2m^2_S}\right)\sin^22\theta
+ \frac{g^2+g^{\prime 2}}{2\lambda^2}\cos^22\theta
-
2\frac{m^2_{H_u}\sin^2\theta + m^2_{H_d}\cos^2\theta -\lambda m^2_0 \sin2\theta}{
m^2_S}.
\eea
The above relation shows that the potential can have at most one local minimum
at $\phi\neq 0$ along the $\phi$-direction for a given $\theta$, which would
appear at $2\lambda^2 \phi^2+m^2_S \sim M^2_{\rm SUSY}$.
On the other hand, along the angular direction, the slope of the potential
vanishes when
\bea
\label{theta-extremum}
\sin2\theta =
-\left(r\frac{\phi^2}{\Lambda^2_1}\sin2\theta
+ \frac{\Lambda^2_2}{\Lambda^2_1} \right)\cos2\theta,
\eea
for which
\bea
\partial^2_\theta V =
4\left(-\cos2\theta+\frac{\Lambda^2_2}{\Lambda^2_1}\sin2\theta \right)
\Lambda^2_1\phi^2 \tan^22\theta.
\eea
Here $r$ is defined by
\bea
r = g^2 + g^{\prime 2} -2\left(1-\frac{A^2_\lambda}{2\lambda^2\phi^2+m^2_S}
\right)\lambda^2,
\eea
and $\Lambda^2_{1,2}$ are given by
\bea
\Lambda^2_1 &=& m^2_{H_d}-m^2_{H_u},
\nonumber \\
\Lambda^2_2 &=& 2\left(1+\frac{A_\lambda B_\kappa}{2\lambda^2\phi^2+m^2_S}
\right)\lambda m^2_0,
\eea
both of which are generally of ${\cal O}(M^2_{\rm SUSY})$, and positive.
For $r>0$, one can find $(i)$ $\partial_\theta V=0$ can have a solution
at $\tan\theta>1$ with a positive curvature $\partial^2_\theta V>0$,
implying that there is only one minimum along the angular direction
for a given $\phi$,
and $(ii)$ for large values of $\phi$, $\phi^2\gg \Lambda^2_1 \sim M^2_{\rm SUSY}$,
a minimum along the angular direction is located near $\tan\theta=1$, i.e.
near the $D$-flat direction.

Let us examine further the case with $r>0$ and $m^2_S\sim M^2_{\rm SUSY}$.
Note that $r$ is positive at $2\lambda^2 \phi^2+m^2_S \sim M^2_{\rm SUSY}$ if $\lambda\lesssim 0.5$ for small $A_\lambda$,
and if $\lambda\lesssim 1$ for $A_\lambda\sim M_{\rm SUSY}$
and $B_\kappa \sim M_{\rm SUSY}$.
At the electroweak vacuum, which lies at $\phi^2 \sim M^2_W \ll M^2_{\rm SUSY}$
and $\theta=\beta$, the condition (\ref{theta-extremum}) gives
\bea
\frac{\Lambda^2_2}{\Lambda^2_1} \simeq -\tan2\beta.
\eea
The extremum condition (\ref{theta-extremum}) can then be written
\bea
\left(
1 - \frac{r\phi^2}{\Lambda^2_1}\frac{\tan^2\theta-1}{\tan^2\theta+1}
\right)\tan2\theta \approx \tan2\beta,
\eea
for $\phi^2\lesssim M^2_{\rm SUSY}$.
This tells that a minimum along the angular direction arises at
$1<\tan\theta<\tan\beta$ for a given $\phi$, and approaches the $D$-flat direction,
$\tan\theta=1$, as $\phi$ increases.
Thus it is useful to first analyze the potential along the $D$-flat direction
though the actual another minimum, if exists, appears somewhat away from
the $D$-flat direction.

On the other hand, in the case with $r<0$, the potential is minimized along the
angular direction at $\tan\theta> \tan\beta$ or at $\tan\theta<1$
for $\phi^2\gtrsim M^2_{\rm SUSY}$.
In the former case, making the potential develop no other minimum in the region
near the $D$-flat direction is not enough to guarantee the absence of a deeper
minimum.

\section{RG running equations}

For the low energy effective theory (\ref{eff-L}) below $M_{\rm SUSY}$,
the RG running equations for the Yukawa couplings read
\bea
8\pi^2\mu \frac{d y^2_t}{d \mu} &=& \left(
\frac{9}{2}y^2_t + y^{\prime 2}_u + y^{\prime 2}_d
- \frac{17}{20}g^2_1 - \frac{9}{4}g^2_2 - 8 g^2_3 \right) y^2_t,
\nonumber \\
8\pi^2\mu \frac{d y^{\prime 2}_u}{d \mu} &=& \left(
3y^2_t + \frac{5}{2} y^{\prime 2}_u + 4y^{\prime 2}_d
- \frac{9}{20}g^2_1 - \frac{9}{4}g^2_2  \right)
y^{\prime 2}_u,
\nonumber \\
8\pi^2\mu \frac{d y^{\prime 2}_d}{d \mu} &=& \left(
3y^2_t + 4 y^{\prime 2}_u + \frac{5}{2}  y^{\prime 2}_d
- \frac{9}{20}g^2_1 - \frac{9}{4}g^2_2 \right)
y^{\prime 2}_d,
\eea
at the one-loop, and those for SM gauge couplings are
\bea
8\pi^2\mu \frac{d }{d \mu}\frac{1}{g^2_1}  &=&
-\frac{103}{30}
-\frac{1}{10}\theta(\mu-m_h)
-\frac{17}{30}\theta(\mu-m_t)
-\frac{2}{5}\theta(\mu-\mu_{\rm eff}),
\nonumber \\
8\pi^2\mu \frac{d }{d \mu}\frac{1}{g^2_2}  &=&
\frac{13}{3} - \frac{1}{6}\theta(\mu-m_h)
-\theta(\mu-m_t)
-\frac{2}{3}\theta(\mu-\mu_{\rm eff}),
\nonumber \\
8\pi^2\mu \frac{d }{d \mu}\frac{1}{g^2_3}  &=&
8 - \theta(\mu-m_t),
\eea
with $g_1=\sqrt{5/3}g^\prime$ and $g_2=g$.
Here $\theta(x)=1$ for $x>0$ and $\theta(x)=0$ for $x<0$.

\end{document}